\documentclass[12pt, letterpaper]{article}
\usepackage{amsmath}
\usepackage{amsfonts}
\usepackage{amssymb}
\usepackage{graphicx}
\usepackage{caption}
\usepackage{subcaption}
\usepackage{graphics,epsfig}
\usepackage{epsf}
\usepackage{epstopdf}
\usepackage{color}

\newcommand{\nc}{\newcommand}
\nc{\ba}{\begin{eqnarray}}
\nc{\ea}{\end{eqnarray}}
\newcommand\be{\begin{equation}}
\newcommand\ee{\end{equation}}

\begin{document}

\begin{flushright} {\footnotesize YITP-19-12, IPMU19-0023}  \end{flushright}
\vspace{5mm}
\vspace{0.5cm}

\def\thefootnote{\fnsymbol{footnote}}

\begin{center}
{\Large {\bf Cosmology in Mimetic SU(2) Gauge Theory}}
\\[0.5cm]
\end{center}

\begin{center}
{ Mohammad Ali Gorji$^{1}\footnote{gorji@ipm.ir}$, Shinji Mukohyama$^{2,3,4}\footnote{shinji.mukohyama@yukawa.kyoto-u.ac.jp}$, Hassan Firouzjahi$^{1}\footnote{firouz@ipm.ir}$ }
\\[.7cm]

{\small \textit{$^{1}$School of Astronomy, 
Institute for Research in Fundamental Sciences (IPM) \\ 
 P.~O.~Box 19395-5531, Tehran, Iran }} \\

 {\small \textit{$^{2}$Center for Gravitational Physics,
 Yukawa Institute for Theoretical Physics \\
 Kyoto University, 606-8502, Kyoto, Japan}} \\

 {\small \textit{$^{3}$Kavli Institute for the Physics and Mathematics of the Universe (WPI), 
 The University of Tokyo Institutes for Advanced Study, 
 The University of Tokyo, Kashiwa, Chiba 277-8583, Japan}} \\

 {\small \textit{$^{4}$Institut Denis Poisson,
 UMR - CNRS 7013, Universit\'{e} de Tours \\
 Parc de Grandmont, 37200 Tours, France}} \\

\vspace{0.5cm}

\end{center}

\vspace{.8cm}

\hrule \vspace{0.3cm}

\begin{abstract} 

It is well known that the standard scalar field mimetic cosmology provides a dark matter-like energy density component. Considering $SU(2)$ gauge symmetry, we study the gauge field extension of the mimetic scenario in spatially flat and curved FLRW spacetimes. Because of  the  mimetic constraint, the standard Yang-Mills term plays the role of the cosmological constant while the mimetic term provides two different contributions: one is the standard radiation scaling like $ a^{-4}$ while the other contribution in  energy density scales  as $\propto a^{-2}$. Consequently, in the Friedmann equation we have two different energy densities which scale as $\propto a^{-2}$: one is the mimetic spatial curvature-like and the other is the standard spatial curvature which can compete with each other. The degeneracy between these two contributions  are disentangled in this scenario since the mimetic spatial curvature-like term shows up only at the dynamical level while the standard spatial curvature term shows up at both dynamical and kinematical levels. 

\end{abstract}

\vspace{0.5cm} 
\hrule
\def\thefootnote{\arabic{footnote}}
\setcounter{footnote}{0}

\newpage

\section{Introduction}

The Einstein theory of general relativity (GR) is not invariant under a conformal transformation. The direct way to implement the conformal symmetry is to construct an action by means of the Weyl tensor which, contrary to the Riemann tensor, is invariant under conformal transformations. These models, however, lead to higher order equations of motion while the standard Einstein equations are second order. For instance, the most simple case is the so-called Weyl-squared gravity which provides fourth-order equations of motion \cite{Weyl:1918ib}. Recently, the mimetic gravity has been suggested in which the conformal degree of freedom of gravity is encoded in an extra scalar field which corresponds to the longitudinal mode of gravity. The model can be realized from the conformal transformation \cite{Mimetic-2013}
\begin{eqnarray}\label{mimetic-trans}
g_{\mu\nu}=( -\tilde{g}^{\alpha\beta} \partial_{\alpha}\phi\partial_{\beta}\phi )\, 
{\tilde g}_{\mu\nu} \, ,
\end{eqnarray}
where $g_{\mu\nu}$ is the physical metric, ${\tilde g}_{\mu\nu}$ is an auxiliary metric and $\phi$ is a scalar field which contains the conformal degree of freedom of gravity which is often called a mimetic scalar. From the above transformation, we see that the physical metric is invariant under the conformal transformation of the auxiliary metric. In comparison with the conformal gravity models that are based on the action containing Weyl tensor leading to higher order equations of motion, the mimetic gravity has second-order equations of motion much similar to GR. Indeed, the transformation Eq. (\ref{mimetic-trans}) is the singular limit of conformal transformations or more general transformations known as disformal transformations \cite{Bekenstein:1992pj}. The number of degrees of freedom does not change under a non-singular conformal/disformal transformation \cite{Domenech:2015tca} while it increases in the case of mimetic transformation Eq. (\ref{mimetic-trans}) so that the scalar field $\phi$ makes the longitudinal mode of gravity dynamical even in the absence of any matter field. In the absence of any matter field, the mimetic term induces energy density scaling  like dark matter and this is the reason why the setup is known as mimetic dark matter.

The mimetic dark matter scenario, however, suffers from caustics formations, beyond which the mimetic scalar is ill-defined. This flaw stems from the mimetic constraint $g^{\mu\nu}\partial_{\mu}\phi\partial_{\nu}\phi=-1$ and the well-known fact that a congruence of timelike geodesics in the influence of gravity forms caustics within the free-fall time. Indeed, the mimetic constraint implies that the hypersurface-orthogonal vector $u^{\mu} \equiv g^{\mu\nu}\partial_{\nu}\phi$ follows the geodesic equation $u^{\nu} \nabla_{\nu} u^{\mu} = 0$ and thus forms caustics, irrespective of the equation of state and the sound speed of the effective fluid. In a conventional cold dark matter scenario, the formation of caustics simply implies that the fluid approximation breaks down and that the system should be described as a collection of particles via the Boltzmann equation. On the other hand, in the mimetic scenario there is no known alternative description when the scalar field description breaks down. Moreover, the sound speed of the curvature perturbations vanishes in the mimetic dark matter scenario and quantum fluctuations of the scalar field are not well-defined. Adding higher derivative term can generate nonzero sound speed \cite{Chamseddine:2014vna} while ghost/gradient instabilities arise \cite{Ramazanov:2016xhp}. Although it is shown that these instabilities can be removed by adding a non-minimal coupling between the scalar field and curvature \cite{Hirano:2017zox}, the resultant models are very complicated \cite{Gorji:2017cai}. It is therefore worthwhile considering modification of the scalar mimetic theory that may potentially avoid caustics formations and also provides healthy propagating scalar mode. 

One such possibility may be to consider a vector field instead of a scalar field, since odd-spin fields may easily exhibit repulsive forces. In Ref. \cite{Gorji:2018okn} (see also Refs. \cite{Barvinsky:2013mea}), as a first step towards such considerations, we have proposed a stable gauge field extension of the mimetic gravity in which the scalar field is replaced by a gauge field so that the conformal transformation Eq. (\ref{mimetic-trans}) is replaced by
\begin{eqnarray}\label{mimetic-trans-gf}
g_{\mu\nu}=\left(-\tilde{g}^{\rho\alpha} \tilde{g}^{\sigma\beta} F_{\alpha\beta} F_{\rho\sigma}
\right)^{\frac{1}{2}}\, {\tilde g}_{\mu\nu} \, ,
\end{eqnarray}
where $F_{\mu\nu} = \partial_{\mu}A_{\nu} - \partial_{\nu}A_{\mu}$ is the field strength of the $U(1)$ gauge field $A_{\mu}$. In comparison with the standard scalar mimetic transformation (\ref{mimetic-trans}), the square root of the kinetic term $F^2$ has appeared in Eq. (\ref{mimetic-trans-gf}). This can be understood, if we note that the inverse of the metric enters twice in $\tilde{g}^{\rho\alpha} \tilde{g}^{\sigma\beta} F_{\alpha\beta} F_{\rho\sigma}$ while it appears only once in the case of scalar mimetic transformation (\ref{mimetic-trans}). It is easy to show that the transformation (\ref{mimetic-trans-gf}) implies the following constraint
\begin{eqnarray}\label{mimetic-const.0}
F^{\mu\nu} F_{\mu\nu} + 1 = 0 \, .
\end{eqnarray}

The gauge fields can have any internal symmetry while the model is defined in the spirit of the spacetime transformation so that the physical metric is invariant under a conformal transformation of the auxiliary metric. In order to study the cosmological implications of the model, in Ref. \cite{Gorji:2018okn} we have considered three $U(1)$ gauge fields and assumed a global $O(3)$ symmetry in the field space, and we have then found the associated cosmological solutions in spatially flat FLRW spacetime. At the background level, we found that the mimetic term with the internal $O(3)$ symmetry induces an energy density component scales as $\propto a^{-2}$ similar to the spatial curvature energy density. This similarity is only held at the dynamical level. The standard spatial curvature, however, not only contributes to the dynamics through the Einstein equations but also changes the geometrical properties of the universe that affects e.g. the geodesic motion. It is therefore natural to explore the cosmological solutions of the gauge field mimetic model in spatially curved FLRW background. This is our main purpose in the present paper. Since in closed and open universes the spatial sector of spacetime have nontrivial closed ${\bf S}^3$ and open ${\bf H}^3$ topologies, we have to consider a complete $SU(2)$ gauge symmetry for the internal field space rather than the combination of the $U(1) \times U(1) \times U(1)$ gauge symmetry and the global $O(3)$ symmetry which we have already studied in Ref. \cite{Gorji:2018okn,Golovnev:2008cf}. The global $O(3)$ symmetry can be considered as the global limit of the $SU(2)$ gauge symmetry and therefore we will have all the results of the case of global $O(3)$ symmetry plus some new effects coming from the local properties of the $SU(2)$ model. We will show that, as we expected, there are two different types of energy density components which behave like $\propto a^{-2}$: the standard one coming from the spatial curvature of the  metric and another one coming from the mimetic term. The first type contributes to both geometry and the dynamics of the universe while the latter contributes only to the dynamics. Therefore, the degeneracy between them breaks in the sense that the geometrical and dynamical spatial curvatures behave differently and  their contributions  in cosmological observables 
can be constrained in different manners by various cosmological observations \cite{Zolnierowski:2015ima}. 

\section{The Model}

Our model is the $SU(2)$ Einstein-Yang-Mills system which in addition respects the mimetic gauge field constraint Eq. (\ref{mimetic-const.0}). The constraint (\ref{mimetic-const.0}) can be taken into account through a Lagrange multiplier \cite{Golovnev:2013jxa} and the action of the model is thus given by
\begin{equation}\label{action0}
S=\int d^{4}x \sqrt{-g}\Big[\, \frac{1}{2} R 
- \lambda' \big( 2 \, {\rm tr}(F_{\mu\nu} F^{\mu\nu}) + 1 \big) 
+ 2 \Lambda\, {\rm tr}(F_{\mu\nu} F^{\mu\nu}) \,\Big] \,,
\end{equation}
where $F_{\mu\nu}=\partial_{\mu} A_{\nu} - \partial_{\nu} A_{\mu} - ig\, [ A_{\mu} , A_{\nu} ]$ is the $SU(2)$ field strength with $g$ is the gauge coupling constant and $\mu,\nu=0,1,2,3$ represent spacetime indices. In our analysis we set $M_P=1$ where $M_P$ is the reduced Planck mass. 

The auxiliary field $\lambda'$  enforces the mimetic constraint
\ba\label{mimetic-const0}
2\,{\rm tr}(F_{\mu\nu} F^{\mu\nu}) = - 1 \,,
\ea
which is the non-Abelian extension of Eq. (\ref{mimetic-const.0}). Note that this is a trivial generalization since the conformal invariance of the mimetic models are defined in the spirit of the spacetime transformations \cite{Gorji:2018okn}.

The constant parameter $\Lambda$ plays the roles of the cosmological constant thanks to the mimetic constraint (\ref{mimetic-const0}). This can be easily seen if we perform the field redefinition $\lambda'\to\lambda=\lambda'-\Lambda$ which transforms the action (\ref{action0}) into the following  form
\begin{equation}\label{action-CC}
S = \int d^{4}x \sqrt{-g}\Big[\, \frac{1}{2} R 
- \lambda \big( 2\,{\rm tr}(F_{\mu\nu} F^{\mu\nu}) + 1 \big) 
- \Lambda\, \,\Big] \,.
\end{equation}

From the above action it is clear that $\Lambda$ is nothing but the cosmological constant term. Since we are interested in the effects of the mimetic term, from now on we neglect the Maxwell term in the action (\ref{action0}) and work with the following action
\begin{equation}\label{action}
S=\int d^{4}x \sqrt{-g}\Big[\, \frac{1}{2} R 
- \lambda \big( 2\,{\rm tr}(F_{\mu\nu} F^{\mu\nu}) + 1 \big) \,\Big] \,,
\end{equation}
which includes only the Einstein-Hilbert term and the mimetic term. The effects of the Maxwell term can be easily taken into account by adding a cosmological constant to the setup.

Now, let us expand the gauge field as $A_{\mu}=A^a_{\mu} T_a$ in which $T_a$ are the generators of the $SU(2)$ group satisfying $[ T_a , T_b ] = i \epsilon_{abc} T_c$ with $a,b,c=1,2,3$ label the gauge group indices. The field strength then takes the following components
\be\label{FST}
F^a_{\mu\nu} = \partial_{\mu} A^a_{\nu} - \partial_{\nu} A^a_{\mu} 
+ g\, \epsilon_{abc} A^b_{\mu} A^c_{\nu} \,,
\ee
where the last term includes the local gauge symmetry effects. In the global limit $g\to0$, the above definition coincides with its global $O(3)$ counterpart which was studied in  \cite{Gorji:2018okn}.

In the component form, the mimetic constraint (\ref{mimetic-const0}) then turns out to be
\ba\label{mimetic-const.}
\delta_{ab} g^{\mu\alpha} g^{\nu\beta} F^{a}_{\alpha\beta} F^{b}_{\mu\nu} = - 1 \,,
\ea
where we have used the fact that ${\rm tr}(T_a T_b) = \frac{1}{2} \delta_{ab}$.

Varying the action (\ref{action}) with respect to the metric, we obtain the Einstein equations
\be\label{EE}
G^\mu_\nu = T^\mu_\nu \,,
\ee
where $G^\mu_\nu $ is the Einstein tensor and the energy-momentum tensor is given by
\be\label{EMT}
T^\mu_\nu = 4\lambda F_a^{\mu\alpha} F^{a}_{\nu\alpha} \,.
\ee
In obtaining the above energy-momentum tensor, we have used the mimetic constraint  (\ref{mimetic-const.}).

Varying the action (\ref{action}) with respect to the gauge field $A^{a}_{\mu}$, we obtain the associated Yang-Mills equations 
\be\label{Maxwell-Eqs}
\nabla_{\mu}F^{a\mu\nu} + g \, \epsilon_{abc} A^b_{\mu} F^{c \mu\nu} 
= - \lambda^{-1}\nabla_{\mu}\lambda\, F^{a\mu\nu} \,.
\ee
The Einstein equations (\ref{EE}) and Yang-Mills equations (\ref{Maxwell-Eqs}) determine the evolution of the metric and gauge fields. In the following sections, we solve these coupled system of equations for the spatially flat and curved FLRW background geometries.

\section{Spatially Flat FLRW}

Before considering the mimetic $SU(2)$ gauge theory with the action Eq. (\ref{action}) in spatially curved FLRW spacetimes, it is useful to consider it in a spatially flat case with the background metric 
\be\label{FLRW-flat-cartesian}
ds^2 = a^2(\tau) \big( - d\tau^2+dx^2+dy^2+dz^2 \, \big) \,,
\ee
where $a(\tau)$ is the scale factor, $\tau$ is the conformal time and $x^i$ are spatial cartesian coordinates. We need an ansatz for the gauge field which is consistent with the symmetry of the background (\ref{FLRW-flat-cartesian}). A consistent ansatz in cartesian coordinates is \cite{Golovnev:2008cf}
\be\label{A-ansatz-cartesian}
g A^a_{\mu} = A(\tau) \delta^a_\mu \,.
\ee
If one being interested only in spatially flat FLRW solutions, it is easier to work in cartesian coordinates. Our aim is, however, to study the model in a spatially curved FLRW background and, therefore, it is better to work with spherical coordinates in which the metric (\ref{FLRW-flat-cartesian}) takes the following form
\be\label{FLRW-flat-spherical}
ds^2 = a(\tau)^2 \big( -d\tau^2 + dr^2 + r^2 (d\theta^2+\sin^2\theta d\varphi^2)\big) \,.
\ee
The gauge field ansatz (\ref{A-ansatz-cartesian}) in terms of spherical coordinates is given by
\begin{eqnarray}\label{A-ansatz}
g A^1_{\mu} &=& A(\tau) 
\big( 0 , \cos\varphi \sin\theta , r\cos\varphi\cos\theta , -r\sin\varphi\sin\theta \big) \,, 
\\ \nonumber
g A^2_{\mu} &=& A(\tau) 
\big( 0 , \sin\varphi \sin\theta , r\sin\varphi\cos\theta , r\cos\varphi\sin\theta \big) \,, 
\\ \nonumber
g A^3_{\mu} &=& A(\tau) 
\big( 0 , \cos\theta , -r\sin\theta , 0 \big) \,.
\end{eqnarray}
Of course, all the results are the same if one works with either Eqs. (\ref{FLRW-flat-cartesian})-(\ref{A-ansatz-cartesian}) or Eqs. (\ref{FLRW-flat-spherical})-(\ref{A-ansatz}). Nonetheless, from the latter expression, we can obtain better intuition about the gauge field ansatz in the more complicated case of the spatially curved FLRW background.

The mimetic constraint Eq. (\ref{mimetic-const.}) implies
\be\label{A-EoM-mimetic-flat}
\frac{6\dot{A}^2}{g^2 a^4} - \frac{6 A^4}{g^2 a^4} = 1 \,,
\ee
where a dot denotes derivative with respect to the conformal time. Solving the above constraint equation gives 
\begin{eqnarray}\label{Adot}
{\dot A} = \mp \sqrt{ A^4 + \frac{g^2 a^4}{6} } \,.
\end{eqnarray}

The Einstein equations (\ref{EE}) give
\begin{eqnarray}\label{Friedmann-flat}
3 {\cal H}^2 = \rho a^2\,, \hspace{1cm} 2\dot{\cal H} + {\cal H}^2 = - p a^2\,,
\end{eqnarray}
where ${\cal H}= \dot{a}/a$ denotes the comoving Hubble expansion rate and we have defined the energy density $\rho = - T^\tau_\tau$ and pressure $p = \frac{1}{3} (T^r_r + T^\theta_\theta + T^\varphi_\varphi)$ as
\begin{eqnarray}
 \rho & = & 2\lambda + 12 \lambda \frac{A^4}{g^2 a^4}  \,, \label{rho-flat}\\
 p & = & - \frac{2\lambda}{3} + 4 \lambda \frac{A^4}{g^2 a^4}  \,.  \label{p-flat}
\end{eqnarray}

The equation of motion for the gauge fields (\ref{Maxwell-Eqs}) implies
\be\label{A-EoM-flat}
\ddot{A} + 2 A^3 = - \frac{\dot{\lambda}}{\lambda} \dot{A}\,.
\ee

In the global limit, i.e. $g\rightarrow0$ with $A/g$ kept constant, Eq. (\ref{Friedmann-flat}) correctly reduces to equations (25) and (26) of Ref. \cite{Gorji:2018okn}. The first terms in (\ref{rho-flat}) and (\ref{p-flat}) containing $\lambda$  come from the global sector of $SU(2)$ gauge symmetry which scale like the spatial curvature energy density with equation of state parameter $w=-1/3$. The second terms, which have the radiation-like equation of state parameter $w = 1/3$, come from the local sector which were absent in \cite{Gorji:2018okn}.

\subsection{Solving background equations}
In order to solve the background equations, it is better to work with the following dimensionless energy density parameters
\be\label{Omega-ED}
\Omega_{\lambda} \equiv \frac{2\lambda}{3H^2} \,, \hspace{1cm}
\Omega_{r} \equiv  \frac{4\lambda}{H^2 g^2}  \frac{A^4}{a^4} \,, \hspace{1cm}
\ee
where $H = a^{-1} {\cal H}$ is the Hubble parameter. The Friedmann equation (\ref{Friedmann-flat}) reduces to
\be\label{Friedmann-flat-Omega}
\Omega_{\lambda} + \Omega_{r} = 1 \,.
\ee

The total energy content of the universe is determined by the radiation $\Omega_r$ which comes from the local sector of the $SU(2)$ gauge symmetry, and $\Omega_{\lambda}$ which is originated from the global sector of $SU(2)$ gauge symmetry and behaves like the spatial curvature energy density. The latter term was also present in  \cite{Gorji:2018okn} where the internal field space had the global $O(3)$ symmetry. Since the contribution of $\Omega_\lambda$ mimics the effects of spatial curvature in Friedmann equation, we denote this term by the ``mimetic spatial curvature-like'' term.

In terms of the energy density parameters defined in Eq. (\ref{Omega-ED}), the Raychuadhuri equation (\ref{Friedmann-flat}) can be written as
\be\label{Hubble-flat}
\frac{1}{H}\frac{dH}{dN} = - (1+\Omega_r) \,,
\ee
where $N=\ln{a}$ is the number of e-folds and also we have used (\ref{Friedmann-flat-Omega}). 

The mimetic constraint (\ref{Adot}) then implies 
\be\label{Mimetic-flat}
\frac{1}{\bar A} \frac{d{\bar{A}}}{dN} = - 1 \mp \frac{1}{\beta H {\bar A}} \sqrt{1+\bar{A}^4} \,,
\ee
where we have defined
\be\label{beta-A0bar}
\bar{A} \equiv \beta \frac{A}{a} \,, \hspace{1cm} \mbox{with} \hspace{1cm} 
\beta \equiv \Big(\frac{6}{g^2}\Big)^{\frac{1}{4}} \,.
\ee
From Eq. (\ref{Omega-ED}), this new variable can be expressed in terms of the dimensionless energy densities as 
\be\label{Abar-ED}
\bar{A}=\Big(\frac{\Omega_r}{\Omega_{\lambda}}\Big)^{\frac{1}{4}}
=\Big(\frac{\Omega_r}{1-\Omega_r}\Big)^{\frac{1}{4}} \,.
\ee
The gauge field equations then yield
\be\label{Maxwell-flat}
\frac{1}{\lambda} \frac{d\lambda}{dN} = - \frac{2}{1+\bar{A}^4} 
\pm \frac{4\bar{A}^3}{\beta H} \frac{1}{\sqrt{1+\bar{A}^4}}\,.
\ee

Differentiating $\Omega_r$ defined in Eq. (\ref{Omega-ED}) and then substituting Eqs. (\ref{Hubble-flat}), (\ref{Mimetic-flat}) and (\ref{Maxwell-flat}) we obtain
\be\label{Omegar-flat}
\frac{1}{\Omega_r} \frac{d\Omega_r}{dN} = - 4 (1-\Omega_r) \mp \frac{4}{\beta H} 
(1-\Omega_r) \big(\Omega_r(1-\Omega_r)\big)^{-\frac{1}{4}} \,,
\ee
where the upper and lower solutions respectively correspond to the the upper and lower solutions in 
Eq. (\ref{Mimetic-flat}).

\begin{figure}
	\caption{Evolution of the energy density parameters in spatially flat universe $k=0$}
	\begin{subfigure}[b]{0.47\textwidth}
		\includegraphics[width=\textwidth]{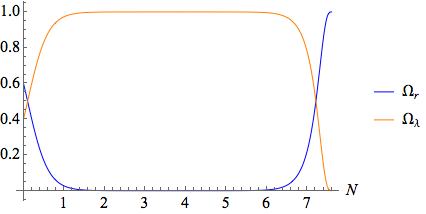}
		\caption{The lower solution of Eq. (\ref{Omegar-flat}):
			The initial conditions are $\Omega_r=0.6$ and $H=10^{-3}$ with 
			the gauge coupling parameter $\beta=10^6$.}
		\label{fig:p-flat}
	\end{subfigure}
	\hfill
	\begin{subfigure}[b]{0.47\textwidth}
		\includegraphics[width=\textwidth]{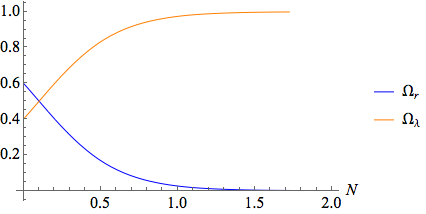}
		\caption{The upper solution of Eq. (\ref{Omegar-flat}):
			The initial conditions are $\Omega_r=0.6$ and $H=10^{-3}$ with 
			the gauge coupling parameter  $\beta=10^6$.}
		\label{fig:n-flat}
	\end{subfigure}
\end{figure}

Now, our task is to find the solution for the system of the first order differential equations (\ref{Hubble-flat}) and (\ref{Omegar-flat}). These equations cannot be solved analytically and therefore we have implemented the numerical methods. The results for two possible cases in Eq. (\ref{Omegar-flat}) are shown in Figs. \ref{fig:p-flat} and \ref{fig:n-flat}.
From Fig.  \ref{fig:p-flat} we see that there are two phases. At first $\Omega_{\lambda}$ dominates while after sufficiently long time $\Omega_r$ finally dominates. This shows that in this case, there is an unstable phase during the radiation dominated era that the mimetic spatial curvature-like energy density dominates and then decays.

Of more interest is the behaviour in Fig. \ref{fig:n-flat} which shows that as time passes, the density parameter $\Omega_{\lambda}$ of the mimetic spatial curvature-like term starts to increase while that of the radiation decreases and finally the mimetic spatial curvature-like energy density dominates. In this case, we have an energy density which behaves like the spatial curvature which appears only at the dynamical level. This type of the energy content is absent in GR. Therefore, it is quite reasonable to study the model (\ref{action})  in spatially curved spacetimes. We therefore have two energy density components with completely different origins which behave in the same manner in the FLRW background at the dynamical level. Since the mimetic spatial curvature-like energy density appears only at the dynamical level, in principle we should  be able to discriminate these energy densities at the geometrical/kinematical level.

\section{Spatially Curved FLRW}

In this section, we find cosmological solutions of the theory described by the action Eq. (\ref{action}) in spatially curved FLRW backgrounds. The metric is given by
\be\label{FLRW-curved}
ds^2 = a(\tau)^2 \big( -d\tau^2 + dr^2 + b(r)^2 (d\theta^2 + \sin\theta^2 d\varphi^2) \big) \,,
\ee
where $b(r) = \sin{r}$ and $b(r) = \sinh{r}$ for the closed ($k=1$) and open ($k=-1$) cases respectively. In the case of $b(r) = r$ we obtain the flat FLRW background ($k=0$) given in Eq. (\ref{FLRW-flat-spherical}).

The line element Eq. (\ref{FLRW-curved}) is invariant under the action of the $G_6$ isometry group which includes spatial rotation and translations. We therefore should consider an ansatz for the 
gauge field which has the same symmetry (up to the $SU(2)$ gauge symmetry). The ansatz Eq. (\ref{A-ansatz}) respects the symmetry of (\ref{FLRW-curved}) only for the flat FLRW case ($k=0$). In order to be able to accommodate the spatially curved ($k=\pm 1$) FLRW backgrounds, we therefore consider the generalization of the Witten ansatz \cite{Witten:1976ck}, following \cite{Galtsov:1991un}. In coordinates defined by the metric (\ref{FLRW-curved}), it is given by
\begin{eqnarray}\label{Witten-Ansatz}
&&g A^a_0 = \omega_0 L^a_1 \,, \hspace{3.35cm} g A^a_1 = \omega_1 L^a_2 \,, \\
&&g A^a_2 = K_2 L^a_2 - (1-K_1) L^a_3 \,, \hspace{1cm}
g A^a_3 = \left[ (1- K_1) L^a_2 + K_2 L^a_3 \right] \sin\theta \,, \nonumber
\end{eqnarray}
where $\omega_0, \omega_1, K_1$, and $K_2$ are functions of $\tau$ and $r$, and $L^a_b$ are defined as
\begin{eqnarray}\label{L_ab}
&& L^a_1 = ( \sin\theta \cos\varphi , \sin\theta \sin\varphi , \cos\theta ) \,, 
\\ \nonumber
&& L^a_2 = ( \cos\theta \cos\varphi , \cos\theta \sin\varphi , - \sin\theta ) \,, 
\\ \nonumber
&& L^a_3 = ( - \sin\varphi , \cos\varphi , 0 ) \,.
\end{eqnarray}

Now, to find the solutions of the unknown functions $\omega_0, \omega_1, K_1$, and $K_2$ in the spatially curved background,  we should insert Eq. (\ref{Witten-Ansatz}) into the Einstein equations with energy-momentum tensor Eq. (\ref{EMT}) and the gauge field equations (\ref{Maxwell-Eqs}). 

Following Ref. \cite{Galtsov:1991un}, we consider the field redefinitions
\begin{eqnarray}\label{new-functions}
\omega_0 \equiv {\bar \omega}_0 + \dot{\alpha} \,, \hspace{1cm}
\omega_1 \equiv {\bar \omega}_1 + \alpha' \,, \hspace{1cm}
K_1 \equiv f \cos\alpha \,, \hspace{1cm}
K_2 \equiv f \sin\alpha \,, 
\end{eqnarray}
where $\alpha,{\bar \omega}_0,{\bar \omega}_1$ and $f$ are functions of $\tau$ and $r$. We remind that a dot represents the derivative with respect to conformal time $\tau$ while a prime indicates a derivative with respect to $r$.  The advantage of working with these new functions is that the function $\alpha$ is pure gauge which does not satisfy any independent equation.

Substituting Eq. (\ref{Witten-Ansatz}) in the mimetic constraint Eq. (\ref{mimetic-const.}) and then using Eq. (\ref{new-functions}), we find the following expression for the mimetic constraint in the spatially curved FLRW background (\ref{FLRW-curved}), 
\be\label{mimetic-constraint}
\frac{2}{g^2 a^4 b^4} 
\Big[ ( f^2 - 1 )^2 - 2 b^2 ( \dot{f}^2 - f'^2 ) - 2 f^2 b^2 ( {\bar \omega}_0^2 - \bar\omega_1^2 )
- b^4 ({\bar\omega}'_0 - \dot{\bar\omega}_1)^2 \Big] = -1 \,.
\ee

Substituting the ansatz (\ref{Witten-Ansatz}) into the energy-momentum tensor Eq. (\ref{EMT}), the nonzero components of the energy-momentum tensor are
\begin{eqnarray}
\label{EMT-00}
T^\tau_\tau &=& - \frac{4\lambda }{ g^2 a^4 b^2 } \Big[ 2 f^2 {\bar \omega}_0^2 
+ 2 \dot{f}^2 + b^2 ({\bar\omega}'_0 - \dot{\bar\omega}_1)^2 \Big] \,,
\\ 
\label{EMT-rr}
T^r_r &=& \frac{4\lambda }{ g^2 a^4 b^2 } \Big[ 2 f^2 {\bar\omega}_1^2 
+ 2 f'^2 - b^2 ({\bar\omega}'_0 - \dot{\bar\omega}_1)^2 \Big] \,,
\\
\label{EMT-thth}
T^\theta_\theta &=& T^\varphi_\varphi = 
 \frac{ 4\lambda }{ g^2 a^4 b^4 } \Big[ ( f^2 - 1 )^2 - b^2 ( \dot{f}^2 - f'^2 ) 
- b^2 f^2 ( {\bar\omega}_0^2 - \bar\omega_1^2 ) \Big] \,,
\\
\label{EMT-0r}
T^\tau_r &=& - \frac{ 8\lambda }{ g^2 a^4 b^2 } 
\big( \dot{f} f' + {\bar\omega}_0 {\bar\omega}_1 f^2 \big) \,.
\end{eqnarray}

Having Eqs. (\ref{EMT-00})-(\ref{EMT-0r}) in hand, we can easily find the Einstein equations for our model (\ref{action}) in the spatially curved background Eq. (\ref{FLRW-curved}).

What remain are the gauge field equations (\ref{Maxwell-Eqs}), which after substituting from Eqs. (\ref{FLRW-curved}) and (\ref{Witten-Ansatz}),  yield
\begin{eqnarray}\label{Maxwell-Eq1}
&& \big[ b^2 ({\bar\omega}'_0 - \dot{\bar\omega}_1) \lambda \big]' 
= 2 \lambda f^2 {\bar\omega}_0 \,,
\\
\label{Maxwell-Eq2}
&& \big[ b^2 (\bar{\omega}'_0 - \dot{\bar\omega}_1) \lambda \dot{\big]\,\,} 
= 2 \lambda f^2 {\bar\omega}_1 \,,
\\
\label{Maxwell-Eq3}
&& f'' - \ddot{f} - \frac{f ( f^2 - 1)}{b^2} - ({\bar\omega}_1^2-{\bar\omega}_0^2) f =
\frac{\dot{\lambda}}{\lambda} \dot{f} - \frac{\lambda'}{\lambda} f' \,.
\end{eqnarray}

In general, the auxiliary field can be a function of the conformal time and radial coordinate. In our case, however, $\lambda=\lambda(\tau)$ is a consistent assumption which also simplifies the equations.

The gauge field equations (\ref{Maxwell-Eq1})-(\ref{Maxwell-Eq3}), together with the Einstein equations with energy-momentum tensor components (\ref{EMT-00})-(\ref{EMT-0r}) and the mimetic constraint Eq. (\ref{mimetic-constraint}) are non-linear coupled system of equations for five unknown functions ${\bar \omega}_0,{\bar \omega}_1$, $f$, $\lambda$, and $a$. In order to solve them, we introduce a new variable $A(\tau)$ as
\be\label{Adot-def}
{\bar\omega}'_0-\dot{\bar\omega}_1 \equiv - \dot{A}\,,
\ee 
and consider the following ansatz
\be\label{f}
f = \sqrt{1+ (A^2-k) b^2} \,. 
\ee
It is useful to note that the function $b(r)$ satisfies  $b'=\sqrt{1-kb^2}$  with $k=0,+1,-1$ corresponding to the flat, closed and open cases respectively. The function $A(\tau)$ is  defined such that it reduces to the function $A(\tau)$ in Eq. (\ref{A-ansatz}) in the flat space limit $k\to0$. Moreover, inspired by \cite{Galtsov:1991un}, we take 
\be\label{Omega1}
{\bar\omega}_1 = \frac{b^2A(A^2-k)}{1+b^2(A^2-k)} \,.
\ee
Substituting (\ref{Adot-def})-(\ref{Omega1}) into the gauge field equations (\ref{Maxwell-Eq1}) and (\ref{Maxwell-Eq2}), we find
\be\label{Omega0}
\bar{\omega}_0 = - \frac{b\dot{A}\sqrt{1-kb^2}}{1+b^2(A^2-k)} \,,
\ee
while equation (\ref{Maxwell-Eq3}) then gives
\be\label{A-EoM}
\ddot{A} + 2 A (A^2-k) = - \frac{\dot{\lambda}}{\lambda} \dot{A}\,.
\ee
Substituting Eqs. (\ref{f})-(\ref{Omega0}) in Eq. (\ref{mimetic-constraint}), the mimetic constraint takes the following simple form
\be\label{A-EoM-mimetic}
\frac{6\dot{A}^2}{g^2 a^4} - \frac{6(A^2-k)^2}{g^2 a^4} = 1 \,.
\ee

From Eq. (\ref{EMT-0r}), we see that the off-diagonal component $T^\tau_r$ of the energy-momentum tensor is not zero. Since the background geometry (\ref{FLRW-curved}) respects the isotropy, the Einstein equations implies $T^\tau_r=0$. Substituting Eqs. (\ref{f})-(\ref{Omega0}) into Eq.  (\ref{EMT-0r}), we can easily find $T^\tau_r = 0$ which show that our setup is consistent. 

Substituting Eqs. (\ref{f})-(\ref{Omega0}) into (\ref{EMT-00})-(\ref{EMT-thth}), we find the following simple forms for the energy density and pressure
\be\label{rho-curved}
\rho = 2\lambda + 12\lambda \frac{ (A^2-k)^2}{g^2 a^4} \,,
\ee
\be\label{p-curved}
p = - \frac{2\lambda}{3} + 4\lambda \frac{(A^2-k)^2}{g^2 a^4} \,.
\ee

The above relations correctly reduce to Eqs. (\ref{rho-flat}) and (\ref{p-flat}) in the flat limit $k\to0$. From Eqs. (\ref{rho-curved}) and (\ref{p-curved}), we see that there are two types of matter components: the first term in the right hand sides which is the mimetic spatial curvature-like term while all the effects of the standard spatial curvature on the mimetic matter is to change the radiation component via $A^2\to A^2-k$. The standard spatial curvature also induces the familiar standard energy density in the geometrical part and the Einstein equations then take the form
\begin{eqnarray}\label{Friedmann-c}
3 {\cal H}^2 &=& ( \rho_k + \rho_\lambda + \rho_r ) \, a^2\,, \\
\label{Raychuadhuri-c}
2\dot{\cal H} + {\cal H}^2 &=& - (p_k + p_\lambda + p_r) \, a^2\,,
\end{eqnarray}
where 
\be\label{rho-k}
\rho_k = -\frac{3k}{a^2} \,, \hspace{1cm} \rho_\lambda = 2\lambda \,,
\hspace{1cm} \rho_r = 12\lambda\frac{ (A^2-k)^2}{g^2 a^4} \,,
\ee
and $p_k = -\frac{1}{3} \rho_k$ and $p_\lambda = -\frac{1}{3} \rho_\lambda$.

The simplest qualitative scenario for the above system is that as time passes, the radiation term first dominates and after it is diluted by the expansion of the universe, the mimetic spatial curvature-like matter together with the standard spatial curvature term dominate. However, one cannot discriminate the latter two from one another at the dynamical level. Moreover, we note that $\rho_\lambda$ interacts with the radiation component $\rho_r$ in the early time which makes  the simple picture  described above more complicated.

\subsection{Solving background equations}

The system of coupled differential equations (\ref{A-EoM}), (\ref{A-EoM-mimetic}), (\ref{Friedmann-c}), and (\ref{Raychuadhuri-c}) cannot be solved analytically. Therefore, as in the case of spatially flat background, we should implement the numerical methods. In order to do so, we define the dimensionless energy density parameters as
\be\label{Omega-ED-c}
\Omega_{\lambda} \equiv \frac{2\lambda}{3H^2} \,, \hspace{1cm}
\Omega_{r} \equiv \frac{4\lambda}{H^2 g^2} \frac{(A^2-k)^2}{a^4} \,, \hspace{1cm}
\Omega_{k} \equiv -\frac{k}{a^2H^2} \,.
\ee
The Friedmann equation (\ref{Friedmann-c}) then simplifies to
\be\label{Friedmann-c-Omega}
\Omega_{k} + \Omega_{\lambda} + \Omega_{r} = 1 \,.
\ee
In comparison with the spatially flat case of Eq. (\ref{Friedmann-flat-Omega}) where both density parameters were positive and less than unity, here the standard spatial curvature energy density can take negative and positive values for the closed and open universes respectively.

From the Raychuadhuri equation (\ref{Raychuadhuri-c}), we find the following equation for the evolution of the Hubble expansion rate in terms of the number of e-folds $N$, 
\be\label{Hubble-c}
\frac{1}{H}\frac{dH}{dN} = - (1+\Omega_r) \,,
\ee
which has the same from as its counterpart in the flat case given by Eq. (\ref{Hubble-flat}). The mimetic constraint Eq. (\ref{A-EoM-mimetic}) then can be rewritten in the following more appropriate form
\be\label{Mimetic-c}
\frac{1}{\bar A} \frac{d{\bar{A}}}{dN}
= - 1 \mp \frac{1}{\beta H{\bar A}} \sqrt{1+(\bar{A}^2+\beta^2 H^2 \Omega_k)^2} \,,
\ee
where $\bar{A}$ and $\beta$ are defined in Eq. (\ref{beta-A0bar}). From Eq. (\ref{Omega-ED-c}), this 
new variable can be expressed in terms of the Hubble parameter and the dimensionless 
energy densities as 
\be\label{Abar-ED-c}
\bar{A}=\pm \bigg( \sqrt{\frac{\Omega_r}{1-\Omega_r-\Omega_k}}
- \beta^2 H^2 \Omega_k \bigg)^{\frac{1}{2}} \,,
\ee
which shows that $\bar{A}$ is determined by the density parameters defined in Eq. (\ref{Omega-ED-c}) and the Hubble expansion rate. Moreover, from the equation of motion of gauge fields (\ref{Maxwell-Eq3}) we find
\be\label{Maxwell-c}
\frac{1}{\lambda} \frac{d\lambda}{dN} = 
- \frac{2}{1+(\bar{A}^2+\beta^2H^2\Omega_k)^2} 
\pm \frac{4\bar{A}}{\beta H} \frac{(\bar{A}^2+\beta^2H^2\Omega_k)
}{\sqrt{1+(\bar{A}^2+\beta^2H^2\Omega_k)^2}} \,.
\ee
Differentiating $\Omega_r$ in Eq.  (\ref{Omega-ED-c}) and then substituting from Eqs. (\ref{Hubble-c})-(\ref{Maxwell-c}) we obtain
\begin{eqnarray}\label{Omegar-c}
\frac{1}{\Omega_r} \frac{d\Omega_r}{dN} 
&=& - 4 + \frac{2\Omega_r(2-\Omega_k)}{1-\Omega_k}
\mp 4 \frac{1-\Omega_k-\Omega_r}{\sqrt{\Omega_r(1-\Omega_k)}}
\bigg(\frac{1}{\beta^2H^2}\sqrt{\frac{\Omega_r}{1-\Omega_k-\Omega_r}}
-\Omega_k \bigg)^{\frac{1}{2}} \,,
\end{eqnarray}
where the upper and lower branches correspond respectively to the the upper and lower solutions in Eq. (\ref{Mimetic-c}). For $\Omega_k=0$, the above equation correctly reduces to its flat counterpart Eq. (\ref{Omegar-flat}). Differentiating $\Omega_k$ in Eq. (\ref{Omega-ED-c}) and then using Eq. (\ref{Hubble-c}) we find
\be\label{Omegak}
\frac{1}{\Omega_k} \frac{d\Omega_k}{dN} = 2 \Omega_r \,.
\ee

As discussed before,  there are two types of energy densities which scale like $a^{-2}$: the standard one coming from the spatial curvature of the  metric (\ref{FLRW-curved}) and mimetic spatial curvature-like matter coming from the global sector of the $SU(2)$ gauge field. Correspondingly, we  define the total dynamical energy density parameter via 
\be\label{Omega-k-tot}
\Omega_k^T=\Omega_k + \Omega_{\lambda}  = 1-   \Omega_{r} \,.
\ee
Consequently, in Friedmann equation, the total curvature term at every moment of the evolution of the 
universe is determined by $\Omega_k^T$ while the dynamics of the model determines how $\Omega_k$ and $\Omega_{\lambda}$ individually contribute to the total energy density parameter. 

We have to find the solutions for $\Omega_k$, $\Omega_r$ and  $H$ from the system of the coupled first order differential equations (\ref{Hubble-c}), (\ref{Omegar-c}), and (\ref{Omegak}). 
The numerical solutions for these functions are presented in Figs.  \ref{fig:p-curved-r}, \ref{fig:n-curved-r}, \ref{fig:p-curved-k}, and \ref{fig:n-curved-k} for the case of open universe and in Figs. \ref{fig:p-curved-r-c}, \ref{fig:n-curved-r-c}, \ref{fig:p-curved-k-c}, and \ref{fig:n-curved-k-c} for the case of closed universe. Below we discuss  the results with various initial conditions for different branches of solutions in Eq. (\ref{Omegar-c}).

\subsubsection{Open universe}
\begin{figure}
	\caption{Evolution of the energy density parameters in open universe $k<0$}
	\begin{subfigure}[b]{0.47\textwidth}
		\includegraphics[width=\textwidth]{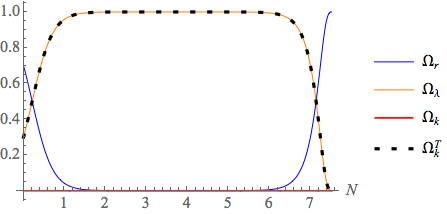}
		\caption{The lower solution of (\ref{Omegar-c}).
		The initial conditions are $\Omega_r=0.7$, $H=10^{-3}$, 
		$\Omega_k=10^{-6}$ with $\beta=10^6$.}
		\label{fig:p-curved-r}
	\end{subfigure}
	\hfill
	\begin{subfigure}[b]{0.47\textwidth}
		\includegraphics[width=\textwidth]{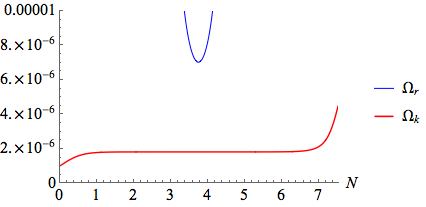}
		\caption{This figure is the same as the left figure but plotted for different ranges of the vertical axis 
		to show the behaviour of $\Omega_k$.}
		\label{fig:p-curved-r2}
	\end{subfigure}
\end{figure}

\begin{figure}
	\caption{Evolution of the energy density parameters in open universe $k<0$}
	\begin{subfigure}[b]{0.47\textwidth}
		\includegraphics[width=\textwidth]{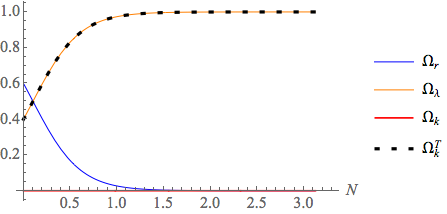}
		\caption{The upper solution of (\ref{Omegar-c}).
		The initial values are $\Omega_r=0.6$, $H=10^{-3}$, and
		$\Omega_k=10^{-6}$ with $\beta=10^6$.}
		\label{fig:n-curved-r}
	\end{subfigure}
	\hfill
	\begin{subfigure}[b]{0.47\textwidth}
		\includegraphics[width=\textwidth]{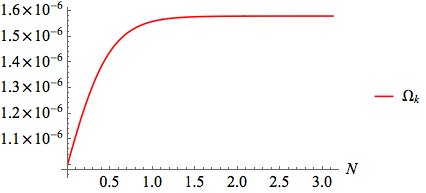}
		\caption{ This figure is the same as the left figure but plotted for different ranges of the vertical axis 
		to show the  behaviour of $\Omega_k$.}
		\label{fig:n-curved-r2}
	\end{subfigure}
\end{figure}

\begin{figure}
	\caption{Evolution of the energy density parameters in open universe $k<0$}
	\begin{subfigure}[b]{0.47\textwidth}
		\includegraphics[width=\textwidth]{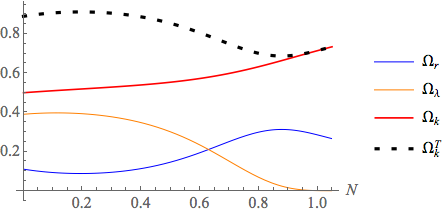}
		\caption{The lower solution of (\ref{Omegar-c}). 
		The figure is plotted for the initial values $\Omega_r=0.11$, $H=10^{-3}$ and 
		$\Omega_k=0.5$ with $\beta=10^2$.}
		\label{fig:p-curved-k}
	\end{subfigure}
	\hfill
	\begin{subfigure}[b]{0.47\textwidth}
		\includegraphics[width=\textwidth]{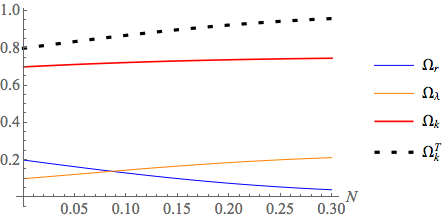}
		\caption{The upper solution of (\ref{Omegar-c}). 
			The initial values are $\Omega_r=0.2$, $H=10^{-3}$ and 
		$\Omega_k=0.7$ with $\beta=10^2$.}
		\label{fig:n-curved-k}
	\end{subfigure}
\end{figure}

For the open universe with $k=-1$, the standard spatial curvature energy density $\Omega_k$, defined in Eq. (\ref{Omega-ED-c}), is always positive and less than unity through the Friedmann equation (\ref{Friedmann-c-Omega}). Plots in  Figs.  \ref{fig:p-curved-r}, \ref{fig:n-curved-r}, \ref{fig:p-curved-k}, and \ref{fig:n-curved-k} show  possible situations for the open universe.

In Fig.  \ref{fig:p-curved-r}, the lower branch of the solution in Eq. (\ref{Omegar-c}) is plotted for the case when the open universe is initially almost radiation dominated $\Omega_r=0.7$ and the standard spatial curvature is very small $\Omega_k=10^{-6}$. The behaviour of the standard spatial curvature is shown in Fig. \ref{fig:p-curved-r2} which shows that it just increases to about $\Omega_k=5\times10^{-6}$. This small amount can be neglected in comparison with the mimetic spatial curvature-like  term which approaches to unity $\Omega_{\lambda}=1$ after about one e-fold. Similar to the lower branch of the flat case in Eq. (\ref{Omegar-flat}), it finally approaches to zero $\Omega_{\lambda}\to0$ and radiation dominates with $\Omega_r\simeq1$. The total energy density parameter $\Omega_k^T$ defined in Eq. (\ref{Omega-k-tot}) is almost equal to the mimetic spatial curvature-like term  most of the time,  $\Omega_k^T\simeq\Omega_{\lambda}$. This is clear from Fig.  \ref{fig:p-curved-r} where the dashed line, representing the total energy density $\Omega_k^T$,  always mimics the solid orange curve which is the mimetic spatial curvature-like term.

The behaviour of the upper solution of Eq. (\ref{Omegar-c}) is plotted in Fig. \ref{fig:n-curved-r}  with the same initial conditions as in Fig. \ref{fig:p-curved-r}. Similar to the previous case  in Fig. \ref{fig:p-curved-r}, the contribution of the standard spatial curvature is always small ( see also Fig.  \ref{fig:n-curved-r2})  
and  most of the time the total spatial curvature is equal to the mimetic spatial curvature-like term. The difference compared to the previous case in Fig. \ref{fig:p-curved-r} is that the universe is finally dominated  by the mimetic spatial curvature-like matter.

The plots in Fig. \ref{fig:p-curved-k} show the lower solution of Eq. (\ref{Omegar-c}) with different initial conditions than the case in Fig. \ref{fig:p-curved-r} in which the universe initially has almost large  spatial curvature $\Omega_k=0.5$ and small fraction of radiation $\Omega_r=0.11$. Therefore, from Eq. (\ref{Friedmann-c-Omega}) we find that initially $\Omega_{\lambda} = 0.39$. As the system evolves, $\Omega_k$ increases and $\Omega_{\lambda}$ decreases so that finally the total spatial curvature is given by the standard spatial curvature,  $\Omega^T_k\simeq\Omega_k$. In this example, the mimetic spatial curvature-like matter has no effects on the final value of the total energy density  $\Omega^T_k$.

The behaviours of the upper solution of Eq. (\ref{Omegar-c}) with somewhat large spatial curvature contribution, $\Omega_k = 0.7$, and small contribution of radiation, $\Omega_r=0.2$, are plotted in Fig.  \ref{fig:n-curved-k}. Interestingly, the contribution of the standard spatial curvature increases slowly while the contribution of the mimetic spatial curvature-like matter increases with a larger rate. The total dynamical spatial curvature $\Omega^T_k$, shown with the dashed curve, then receives  contributions from both the standard spatial curvature and the mimetic spatial curvature-like matter.

\subsubsection{Closed universe}

\begin{figure}
	\caption{Evolution of the energy density parameters in closed universe $k>0$}
	\begin{subfigure}[b]{0.47\textwidth}
		\includegraphics[width=\textwidth]{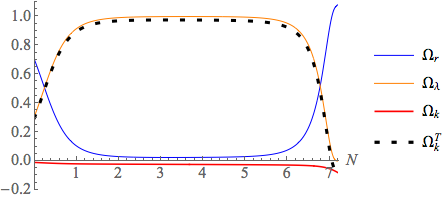}
		\caption{The lower solution of (\ref{Omegar-c}). 
			The initial conditions are $\Omega_r=0.7$, $H=10^{-3}$, 
			$\Omega_k=-10^{-2}$ with $\beta=10^6$.}
		\label{fig:p-curved-r-c}
	\end{subfigure}
	\hfill
	\begin{subfigure}[b]{0.47\textwidth}
		\includegraphics[width=\textwidth]{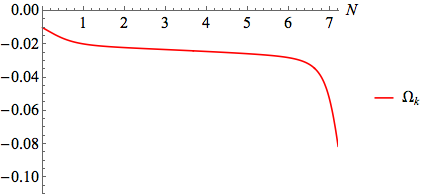}
		\caption{ This figure is the same as the left figure but is focused on different ranges
			to see the  behaviour of  $\Omega_k$.}
		\label{fig:p-curved-r2-c}
	\end{subfigure}
\end{figure}

\begin{figure}
	\caption{Evolution of the energy density parameters in closed universe $k>0$}
	\begin{subfigure}[b]{0.47\textwidth}
		\includegraphics[width=\textwidth]{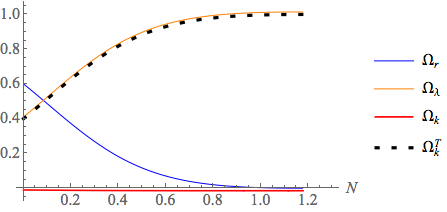}
		\caption{The upper solution of (\ref{Omegar-c}). 
			The initial values are $\Omega_r=0.6$, $H=10^{-3}$, and
			$\Omega_k=-10^{-2}$ with $\beta=10^6$.}
		\label{fig:n-curved-r-c}
	\end{subfigure}
	\hfill
	\begin{subfigure}[b]{0.47\textwidth}
		\includegraphics[width=\textwidth]{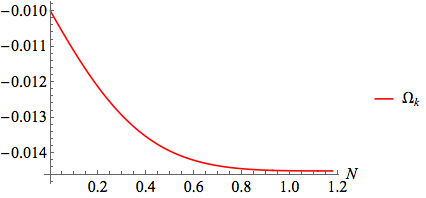}
		\caption{This figure is the same as the left figure but is focused on different ranges
			to see  the behaviour of  $\Omega_k$.}
			\label{fig:n-curved-r2-c}
	\end{subfigure}
\end{figure}

\begin{figure}
	\caption{Evolution of the energy density parameters in closed universe $k>0$}
	\begin{subfigure}[b]{0.47\textwidth}
		\includegraphics[width=\textwidth]{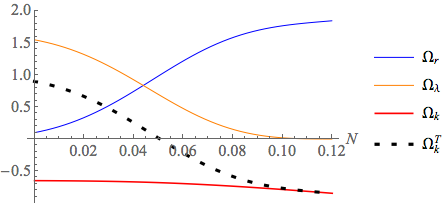}
		\caption{The lower solution of (\ref{Omegar-c}). 
			The figure is plotted for the initial values $\Omega_r=0.1$, $H=10^{-3}$ and 
			$\Omega_k=-0.65$ with $\beta=10^2$.}
		\label{fig:p-curved-k-c}
	\end{subfigure}
	\hfill
	\begin{subfigure}[b]{0.47\textwidth}
		\includegraphics[width=\textwidth]{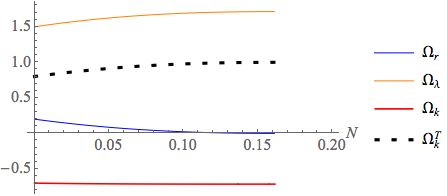}
		\caption{The upper solution of (\ref{Omegar-c}). 
			The initial values are: $\Omega_r=0.2$, $H=10^{-3}$ and 
			$\Omega_k=-0.7$ with $\beta=10^2$.}
		\label{fig:n-curved-k-c}
	\end{subfigure}
\end{figure}

For the closed universe with $k=1$, the density parameter of the standard curvature term is negative, $\Omega_k <0$.  We have plotted the possible situations in Figs.  \ref{fig:p-curved-r-c}, \ref{fig:n-curved-r-c}, \ref{fig:p-curved-k-c}, and \ref{fig:n-curved-k-c} for this case.

The lower solution in Eq. (\ref{Omegar-c}) for a closed universe is plotted in Fig. \ref{fig:p-curved-r-c}. The initial conditions are considered so that the universe is almost radiation dominated with $\Omega_r=0.7$ while the spatial curvature is small $\Omega_k=-10^{-2}$. The result shows that the contribution of the 
total dynamical spatial curvature is almost equal to the contribution of the mimetic spatial curvature-like matter which finally approaches zero. The evolution of the contribution of the standard spatial curvature is shown in figure \ref{fig:p-curved-r2-c} which shows a slight enhancement that is negligible in the total spatial curvature. 

The upper solution of Eq. (\ref{Omegar-c})  for $\Omega_r=0.6$ and $\Omega_k=-10^{-2}$ is plotted in Fig.  \ref{fig:n-curved-r-c}. Similar to the previous case shown in Fig. \ref{fig:p-curved-r-c}, the contribution from the standard spatial curvature is negligible (see Fig.  \ref{fig:n-curved-r2-c}) and the total spatial curvature is almost equal to the mimetic spatial curvature-like term. Contrary to the previous case, however, the mimetic spatial curvature-like matter finally dominates.

In Fig. \ref{fig:p-curved-k-c} the lower solution of Eq. (\ref{Omegar-c}) is plotted for $\Omega_r=0.1$ and $\Omega_k=-0.65$. The contribution of the mimetic spatial curvature-like matter decreases and approaches zero and the total energy density is finally given only by the standard spatial curvature.

The upper solution of Eq. (\ref{Omegar-c}) for $\Omega_r=0.2$ and $\Omega_k=-0.7$ is plotted in Fig.  \ref{fig:n-curved-k-c}. The closed universe ends up with the spatial curvature dominated era so that the total spatial curvature receives positive and negative contributions from the mimetic spatial curvature-like matter and the standard spatial curvature respectively.

There are two interesting situations which can potentially happen in this scenario: i) The mimetic spatial curvature-like matter being the dominant contribution in the total spatial curvature $\Omega^T_k$. We have seen that this is possible in both open and closed universes which is clear from Figs. \ref{fig:n-curved-r} and \ref{fig:n-curved-r-c} for the open and closed universes respectively. We therefore can have significant spatial curvature energy density at the dynamical level while it can be very small at the geometrical level where only the standard spatial curvature is present. This is very different than what we have in GR where spatial curvature of the metric determines the total spatial curvature at both the geometrical and dynamical levels. ii) The mimetic spatial curvature-like matter and the standard spatial curvature can have opposite signs and therefore can cancel each other. In this respect, we can have no significant total spatial curvature at the dynamical level and the model is almost like in GR without any spatial curvature at the background level while it is expected to be completely different at the level of perturbations.

From  above discussions we have seen that the mimetic spatial curvature-like matter $\Omega_{\lambda}$ contributes to the total spatial curvature of the universe $\Omega^T_k$  at the dynamical level and it can compete with the standard spatial curvature $\Omega_k$. So, it is interesting if we could discriminate between the mimetic spatial curvature-like matter and the standard spatial curvature. In order to do this, we note that the mimetic matter only appears at the level of dynamics while the standard spatial curvature appears at both the dynamical and geometrical levels. For instance, the mimetic spatial curvature-like matter has nothing to do with the geodesic equation which determines the path of light in the gravitational lensing scenario while the standard spatial curvature of the metric affects the geodesic equations. Therefore, we can distinguish between the mimetic spatial curvature-like matter and the standard curvature term via two different cosmological observations. 

\section{Summary and Conclusions}

As a unique singular limit of scalar conformal transformations, the mimetic gravity can provide an energy density which mimics the roles of the dark matter in cosmological backgrounds. The conformal mode of the gravity is encoded in the scalar field in this scenario and it is plausible to replace it with any other field to find other possible mimetic scenarios. In the case of a gauge field, in which we are interested in this paper, we have to consider a symmetry homomorphic to the $O(3)$ symmetry for gauge fields in order to be able to find an isotropic cosmological solution. The most simple choice is the $O(3)$ symmetry itself which we have considered in Ref. \cite{Gorji:2018okn}. We have found there that the model provides energy density which behaves as $\propto a^{-2}$ in spatially flat FLRW background. It is then interesting to extend the setup to the case of spatially curved FLRW background and see how the total spatial curvature behaves in the presence of the genuine spatial curvature-like matter. In order to do this, however, we have to take into account the $SU(2)$ gauge symmetry due to the nontrivial topologies of the spatial sectors of the metric in the spatially curved FLRW spacetimes. Therefore, we have studied the mimetic $SU(2)$ gauge theory in this paper. We first studied the case of mimetic $SU(2)$ gauge theory in the spatially flat FLRW background which is useful for the comparison of the results of the spatially curved setup with the spatially flat case. The mimetic constraint forces the kinetic term of the gauge field to be constant so that it behaves like the cosmological constant. The mimetic sector provides two energy density components: one behaving like radiation $\propto a^{-4}$ and another behaving $\propto a^{-2}$. In Friedmann equation, the total dynamical spatial curvature then consists of two different components: one is the standard geometrical component and another  coming from the mimetic term. The latter has only dynamical contributions   and it is absent at the geometrical level while the standard spatial curvature is present at both the dynamical and geometrical levels. In this sense, we found that the degeneracy of the spatial curvature at the dynamical and geometrical levels is resolved in this setup. If we have a non-degenerate observations which separately measure the dynamical and geometrical properties of the universe, we can find observational signatures of the model.

\vspace{01cm}


{\bf Acknowledgments:}  M. A. Gorji thanks F. Hajkarim and M. Sasaki for useful discussions. He also thanks the Yukawa Institute for Theoretical Physics (YITP) at Kyoto University for hospitality during the ``2019 YITP Asian-Pacific Winter School and Workshop on Gravitation and Cosmology" and Kavli Institute for Physics and Mathematics of the Universe (IPMU) at Tokyo University for hospitality where this work was in its final stage. The work of S. Mukohyama was supported by Japan Society for the Promotion of Science (JSPS) Grants-in-Aid for Scientific Research (KAKENHI) No. 17H02890, No. 17H06359, and by World Premier International Research Center Initiative (WPI), MEXT, Japan. H. Firouzjahi would like to thank DAMTP, Cambridge University for hospitality where this work was in its final stage.


{}

\end{document}